\input harvmac

\def \la {\langle}
\def \ra {\rangle}
\def \e {{\rm e}}
\def \ov {\over}

\def \pa { \partial}
\def \a {\alpha}
\def \b {\beta}

\def \G {\Gamma}
\def \d {\delta}
\def \m {\mu}
\def \n {\nu}
\def \s {\sigma}
\def \ee {\epsilon}
\def \r {\rho}

\def \ta {\tau}

\def \ev {\varepsilon}

\def \dx {\dot{x}}
\def \lr { \lref}
\def\np {{  Nucl. Phys. }}

\def \rf {\refs} 

\lr \tsr{ A.A.~Tseytlin, ``Sigma model approach to string theory 
effective actions with tachyons", hep-th/0011033.}

\lr \an{O.D.~Andreev and A.A.~Tseytlin,
``Partition Function Representation For The Open
Superstring Effective Action: Cancellation Of Mobius
Infinities And Derivative Corrections To
Born-Infeld Lagrangian,''
Nucl.\ Phys.\  {\bf B311}, 205 (1988).}

\lr \ts { A.A.~Tseytlin,
``Renormalization Of Mobius Infinities And Partition
Function Representation For String Theory Effective
Action,''
Phys.\ Lett.\  {\bf B202}, 81 (1988).
 }

\lr \wi { E. Witten, ``On background-independent
open-string field theory,'' Phys. Rev. {\bf D46}, 5467
(1992), hep-th/9208027.
``Some computations in background-independent
off-shell string theory,'' Phys. Rev. {\bf D47}, 3405 
(1993),  hep-th/9210065.
}

\lr \oku{ K. Okuyama, ``Noncommutative Tachyon from Background
Independent Open String Field Theory,''
 hep-th/0010028.}

\lr\corn{
L. Cornalba, ``Tachyon condensation in large
magnetic
fields with background independent string field theory,''
hep-th/0010021.} 

\lr \ed {E.~Witten,
``D-branes and K-theory,''
JHEP {\bf 9812}, 019 (1998), 
hep-th/9810188.
}

\lr\km{D. Kutasov, M. Mari\~no, and G. Moore, ``Some
exact results on tachyon condensation in string
field theory,'' hep-th/0009148.}

\lr \sen {A.~Sen,
``Non-BPS states and branes in string theory,''
hep-th/9904207.
}

\lr \gsh{A. Gerasimov and S. Shatashvili, ``On exact tachyon
potential
in open string field theory,'' JHEP {\bf 0010}, 034 (2000), 
hep-th/0009103.
}

\lr \kmm   {D.~Kutasov, M.~Mari\~no and G.~Moore,
``Remarks on tachyon condensation in superstring field
theory,''
hep-th/0010108.
 } 

\lr \sh {S. Shatashvili, ``Comment on the background independent
open string theory,'' Phys.\ Lett.\  {\bf B311}, 83
(1993),
hep-th/9303143;
``On the problems with background independence in string theory,''
hep-th/9311177.}

\lr\ftbi {E.S.~Fradkin and A.A.~Tseytlin,
``Nonlinear Electrodynamics From Quantized Strings,''
Phys.\ Lett.\  {\bf B163}, 123 (1985).
}

\lr\ft { E.S.~Fradkin and A.A.~Tseytlin,
``Quantum String Theory Effective Action,''
Nucl.\ Phys.\  {\bf B261} (1985) 1.
``Effective Field Theory From Quantized Strings,''
Phys.\ Lett.\  {\bf B158}, 316 (1985).
}

\lr\witli {K. Li and E. Witten,
``Role of short distance behavior in off-shell open-string
field theory,'' 
Phys. Rev. {\bf D48}, 853 (1993), hep-th/9303067.}

\lr\and {O.~Andreev,
``Some computations of partition functions and
tachyon potentials in  background independent
off-shell string theory,''
hep-th/0010218.
}

\lr\ttu{
T.~Takayanagi, S.~Terashima and T.~Uesugi,
``Brane-antibrane action from boundary string field theory,''
hep-th/0012210.
}

\lr\krl{
P.~Kraus and F.~Larsen,
``Boundary string field theory of the D D-bar system,''
hep-th/0012198.
}

\lr\aftt{
G.~Arutyunov, S.~Frolov, S.~Theisen and A.~A.~Tseytlin,
``Tachyon condensation and universality of DBI action,''
JHEP{\bf 0102}, 002 (2001)
hep-th/0012080.
}

\lr\gs{
D.~Ghoshal and A.~Sen,
``Normalisation of the background independent open string 
field theory  action,''
JHEP{\bf 0011}, 021 (2000)
hep-th/0009191.
}
\lr\kpp{
V.~A.~Kostelecky, M.~Perry and R.~Potting,
``Off-shell structure of the string sigma model,''
Phys.\ Rev.\ Lett.\ {\bf 84}, 4541 (2000),
hep-th/9912243.
}
\lr\pol{
J.~Liu and J.~Polchinski,
``Renormalization Of The Mobius Volume,''
Phys.\ Lett.\ B {\bf 203}, 39 (1988).
}
\lr\mar{
M.~Marino,
``On the BV formulation of boundary superstring field theory,''
hep-th/0103089.
}
\lr\np{
V.~Niarchos and N.~Prezas,
``Boundary superstring field theory,''
hep-th/0103102.
}
\lr\has{
K.~Hashimoto and S.~Hirano,
``Metamorphosis of tachyon profile in unstable D9-branes,''
hep-th/0102174.
}
\lr\buch{
I.~L.~Buchbinder, V.~A.~Krykhtin and V.~D.~Pershin,
``Massive field dynamics in open bosonic string theory,''
Phys.\ Lett.\ B {\bf 348}, 63 (1995)
[hep-th/9412132].
}
\lr\laba{
J.~M.~Labastida and M.~A.~Vozmediano,
``Bosonic Strings In Background Massive Fields,''
Nucl.\ Phys.\ B {\bf 312} (1989) 308.
}
\lr\foer{
S.~Foerste,
``Generalized conformal invariance conditions on a sigma model of the
open bosonic string including the first massive mode,''
Annalen Phys.\ {\bf 1} (1992) 98.
}

\baselineskip8pt
\Title{\vbox
{\baselineskip 6pt{\hbox{OHSTPY-HEP-T-01-007  }}{\hbox
{   }}{\hbox{hep-th/0104042}} {\hbox{
   }}} }
{\vbox{\centerline {On off-shell structure of open string sigma model }
\medskip
 \centerline { }
 }}
\vskip -20 true pt

\medskip
\centerline{   Sergey A. Frolov \footnote{$^{\star}$}{\baselineskip8pt
e-mail: frolov@mps.ohio-state.edu}
\footnote{$^{\dagger}$}{\baselineskip8pt
Also at Steklov Mathematical Institute, Moscow.}
}

\smallskip\smallskip
\centerline {\it  Department of Physics }
\smallskip
\centerline {\it  The Ohio State University}\smallskip
\centerline {\it  Columbus, OH 43210-1106, USA}

\bigskip\bigskip
\centerline {\bf Abstract}
\medskip
\baselineskip10pt
\noindent
\medskip

We analyze several problems related to off-shell structure of
open string sigma model by using a combination of derivative
expansion and expansion in powers of the fields.
According to the sigma model approach to
bosonic open string theory, the tachyon effective action $S(T)$
coincides with  the renormalized partition function $Z(T)$ of
sigma model on a disk, up to a term vanishing on shell.
On the other hand, $Z(T)$ is a generating functional of perturbative
open string scattering amplitudes. If $S(T) = Z(T)$, then there
should be no contribution of exchange diagrams to string amplitudes
computed using $S(T)$. We compute the cubic term in the effective
action, and show that it vanishes if some but not all external
legs are on shell, and, therefore, any exchange diagram involving
the cubic term vanishes too. Then, we discuss a problem of turning on
nonrenormalizable boundary interactions, corresponding to massive
string modes. We compute the quadratic term for a
symmetric tensor field, and show that despite nonrenormalizability
of the model one can consistently remove all divergent terms, and
obtain a quadratic action reproducing the on-shell condition for
the field. We also briefly discuss fermionic (NS) sigma model,
compute the tachyon quadratic term, and show that it reproduces
the correct  tachyon mass. We note that turning on a massive
symmetric tensor field leads to the appearance of a term linear
in it, which can be removed by adding a higher-derivative
term to the boundary of the disc.

\Date {April  2001}

\noblackbox
\baselineskip 16pt plus 2pt minus 2pt
\newsec{Introduction}
The boundary sigma model approach 
\rf{\ft,\ftbi,\ts,\an,\wi,\witli,\sh}
(for a review, see \tsr) to open string theory has been successfully 
applied to the study of tachyon condensation
\sen\ in open string theory
\rf{\gsh,\km,\gs,\corn,\oku,\kmm,\and,\aftt,\krl,\ttu,\has}.
According to  the approach,  
the effective action $S(T, A)$  for the massless
vector and tachyon fields in
bosonic open string theory is related to  the renormalized partition 
function $Z(T,A)$ of 
boundary sigma model on the disk as \sh
\eqn\sz{
S(T, A)=\biggl(1+\b^T{\pa\over\pa T}+\b^A{\pa\over\pa A}\biggr)Z(T,A),
}
where $\b^T$ and  $\b^A$ are tachyon and vector fields beta-functions, 
respectively.

On the other hand, $Z(T,A)$ is a generating
functional of perturbative open string
scattering amplitudes, and since the
beta-functions vanish on shell, then, naively,
according to \sz, a $n$-point string
scattering amplitude would be given
just by the corresponding term in $S(T,A)$.
This  is a puzzle, because
one also would expect nonvanishing contribution of  exchange
diagrams to string amplitudes computed using $S(T,A)$.
In the case of the exchange by a massless particle
the puzzle was resolved
in \ts\ by noting that the renormalized partition
function $Z(T,A)$ does not generate
string scattering amplitudes because
renormalization of logarithmic infinities
corresponds to subtraction of
massless poles in the amplitudes.
However, this does not explain what
happens with the exchange of tachyons.
In particular, if one
considers the tachyon effective action, i.e. one sets $A=0$,\foot{
It can be done consistently because there is no term
linear in the vector field in
the effective action.} then one should explain
why the tachyon cubic term
does not contribute to the 4-point
tachyon scattering amplitude through
exchange diagrams.

To try to resolve the puzzle,
we compute the tachyon cubic term
in the effective action $S(T)$, and show that
it vanishes if not all external legs are on shell, and, therefore,
any exchange diagram involving the cubic terms vanishes too.
However, if all external legs are on shell,
the cubic term is ill-defined,
and to obtain a well-defined string scattering amplitude one has to
shift the on-shell mass condition by introducing a soft mass term for
the tachyon. We expect that any term
in the tachyon effective action
possesses the same property, and this gives a resolution 
of the problem. Let us note that the same tachyon cubic term was
previously computed in \km, but we are unable to check
that our expression coincides with theirs.

One may try to generalize the boundary sigma model by turning on
nonrenormalizable interactions corresponding to massive string modes,
i.e. to deal with the boundary string field theory (BSFT)
\rf{\wi,\sh}. It is believed that
once one includes a massive string mode
one has to turn on all the string modes.
Although this is so in general,
we argue that one can consistently reconstruct the part of the string
effective action which depends on all string
modes up to some mass level
by defining the action as a series in powers of the modes.
As an example,
we compute quadratic terms for tachyon, vector and symmetric
tensor fields, and show that despite the nonrenormalizability of
the model one can consistently remove all divergent terms, and
obtain a quadratic action reproducing the on-shell conditions for the
fields. The same equations of motion were previously derived from
the conformal invariance condition of the open string sigma model
in \rf{\laba,\foer,\buch}.

We also briefly discuss the fermionic boundary sigma model.
We compute the tachyon quadratic term and show that it exhibits the
required zero at $p^2 = {1\ov 2\a'}$. We then comment on the
inclusion of massive modes in the sigma model.
We note that turning on a massive
symmetric tensor field leads to the appearance of a term linear
in it, which can be removed by adding a proper higher-derivative
term to the disc boundary. It is unclear how this may influence
recent discussions of fermionic BSFT in \rf{\mar,\np}.

The plan of the paper is as follows. In Section 2 we compute the
tachyon cubic term and
show that it does not contribute to scattering amplitudes.
In Section 3 we discuss massive string modes
and compute the quadratic
action for tachyon, vector and symmetric tensor
of the second rank.
In Section 4 we derive the tachyon quadratic
action for the fermionic case, and determine a
higher-derivative term which can be added to the disc boundary to cancel
a term linear in massive symmetric tensor field.
In Appendix A details of
the computation of the tachyon cubic term are presented.
In Appendix B some useful formulas are collected.

\newsec{Tachyon cubic term and exchange diagrams}
The bosonic sigma model with the boundary tachyon interaction
is described by the action
$$
S = S_{\Sigma } + S_{\pa \Sigma },
$$
where
$$
S_{\Sigma } = {1\over 4\pi} 
\int_ {\Sigma }\, d^2\, z\  \pa x^\m \bar{\pa} x_\m ,$$
and
$$
S_{\pa \Sigma } =  \int_0^{2\pi} \, {d\tau 
\over 2\pi}\ {1\ov \ee} T(x(\ta ))
\  .
$$
Here $\Sigma$ is a disc of unit radius, $T$ is the tachyon of
the bosonic open string theory, $\ee$ is a UV cutoff, and $\a' =2$.
To find the tachyon effective action,
we first have to compute the partition function
\eqn\zta{
Z(T)=\la {\rm e}^{-S_{\pa \Sigma }}\ra ,
}
where the averaging is performed by using the free bulk action 
$S_{\Sigma }$.

According to the boundary sigma model approach, the partition function
has to be computed in the framework of the $\a'$-expansion that is
equivalent to the expansion in derivatives of tachyon.
It is usually said that the derivative expansion is incompatible with
the expansion in powers of tachyon. However, this is so only if one
takes tachyon to be near its mass shell\foot{Off-shell
structure of the sigma model in this case
was recently discussed in \kpp.}. The partition function computation
in powers of tachyon is usually done by
representing tachyon in the form
\eqn\Tp{
T(x(\ta ))=\int d^d p\ T(p)\e^{ip \cdot x} \e^{ip\cdot \xi (\ta )} .}
Here $p\cdot  x\equiv p_\mu x^\m$, we denote the zero mode of
$x(\ta )$ as $x$,
and write
$$x(\ta ) = x + \xi (\ta ),\quad \int\, d\, \ta \xi (\ta ) =0.$$
Then it is easy to see that
the derivative expansion just means that one should do all
computations in the vicinity of $p=0$.

The expansion of $Z(T)$ in powers of tachyon
up to the third order is given
by\foot{The partition function and the
effective action have to be multiplied by the constant which
is equal to the value the free theory partition
function on the disc (with all boundary fields turned off)
and coincides with the D25-brane
tension. We omit the multiplier throughout the paper.}
\eqn\ztac{
Z(T,A)= \la 1\ra +T+ TT + TTT ,
}
where 
$$
\la 1\ra = \int d^dx,
$$
\eqn\T{
T= - {1\ov\ee}
\int \, {d\tau\ov 2\pi}\
\la\ T(x(\ta )) \ \ra = - {1\ov\ee}\int\ d^dx\ T(x),
}
\eqn\TT{
TT=   {1\ov 2} {1\ov \ee^2}
\int \, {d\tau\ov 2\pi}{d\ta ' \ov 2\pi}\
\la\ T(x(\ta )) T(x(\ta '))\ \ra ,
}
\eqn\TTT{
TTT= -{1\ov 6}{1\ov \ee^3}
\int \, {d\tau_1\ov 2\pi}{d\ta_2 \ov 2\pi}{d\ta_3 \ov 2\pi}\
\la\ T(x(\ta_1 )) T(x(\ta_2))T(x(\ta_3))\ \ra .}

To compute the terms we use \Tp\ and the boundary
bosonic Green function 
\eqn\bgrff{
G_B^{\m\n}(\tau):= \la \xi^\m(\tau)\xi^\nu(0)\ra = \d^{\m\n}G_B(\tau)
=-\d^{\m\n}\left(  2\log\left( \sin^2\left({\ta\ov 2}\right)+\ee^2
\right) +2\log 4  \right) .
}
The details of the computation  and the expressions  for the terms
can be found in the Appendix A. Here we only comment on the
computation of the quadratic term $TT$. By using \Tp\ and
formula (B.1) from Appendix B one finds the following
behavior of $TT$ at
small $\ee$
$$
TT={1\ov 2}(2\pi )^{d}\int d^dp\ T(p)T(-p)\
\left(
\ee^{4p^2-2} {\Gamma ( {1\ov 2} - 2p^2 )\ov \G ( {1\ov
2} ) \G (1-2p^2)} + {1\ov\ee} {\Gamma ( -{1\ov 2} + 2p^2 )
\ov \G ( {1\ov
2} ) \G (2p^2)} \right) .
$$
It is seen from this expression that for small $p$ the first term
dominates and exhibits the required zero at the tachyon mass shell
$p^2 = {1\ov 2}$.
Renormalization of the quadratic term in the partition function
is done by renormalizing the
bare $T$ as
\eqn\tren{
T(p)\ \ee^{2p^2-1} = T_{R}(p)\ \ee_{{}_R}^{2p^2} .}
It can be also shown that the same renormalization makes any term
$T^n$ in the partition function finite, and, therefore,
the exact tachyon beta-function is equal to
\eqn\tacb{
\b_T (p) = (2p^2-1)T(p).
}
However, if one would interested in $TT$ near
the tachyon mass shell, the second term would dominate and give
a power divergent term, which would make the tachyon beta-function
nonlinear.

The corresponding terms in the effective action can be easily found by
using \sz\ and the tachyon beta-function \tacb.
Here we list and discuss the results obtained
(we omit the subscript $R$
on $T$ in what follows).
\eqn\TTs{
S_{TT}=  -(2\pi )^{d}\int d^dp\ T(p)T(-p)\
\ee_{{}_R}^{4p^2}
{\Gamma ( {3\ov 2} - 2p^2 )\ov \G (  {1\ov 2}  ) \G (1-2p^2)}  .
}
Up to the renormalization factor
$\ee_{{}_R}^{4p^2}$ the
expression coincides with the one obtained in \km. It is clear that
the integrand vanishes at $ p^2 =  {1\ov 2} $,
that is the correct tachyon mass.
Expanding the tachyon quadratic term in powers of $p^2$, what is
equivalent to the derivative expansion of
the effective action, we get
the term in the two-derivative approximation 
$$
S_{TT}=\int d^dx\ \left( - {1\ov 2} T^2 +
(2 - \log (4\ee_{{}_R}^2))\pa_\m T
\pa^\m T\right) .
$$ 
The first term $ - {1\ov 2} T^2$ has the correct
coefficient that follows from
the potential $(1+T){\rm e}^{-T}$. There are
three interesting choices
of $\ee_{{}_R}$: $i)$ $\ee_{{}_R}^2  =
{{\rm e}\ov 4}$ gives the correct
tachyon mass already at
the two-derivative approximation.
$ii)$ $\ee_{{}_R}  = {1\ov 2}$ leads to the
two-derivative approximation of the
action obtained in \km.  This choice of
$\ee_{{}_R}$ seems to be the
most convenient to study the tachyon condensation.
$iii)$ $\ee_{{}_R}  = {{\rm e}\ov 2}$. Under this choice of $\ee$
there is no two-derivative
tachyon term in the effective action.

It is worth noting that the integrand in \TTs, in fact, exhibits
infinite number of zeroes and poles at $2p^2= 1 + n$ and
$2p^2 = {3\ov 2}+ n$, respectively. This probably indicates that
the expansion in powers of $T$ is well-defined only in some region
of $p^2$ which includes, however, the tachyon mass shell.

The tachyon cubic term in the effective action \sz\ is given by
\eqn\TTTs{
S_{TTT}= {(2\pi )^d\ov 3}\int d^dp_i\ \d (\sum_i p_i)\
T(p_1 ) T(p_2)T(p_3)\
\left( \ee_{{}_R}^2\right)^{\sum_i p_i^2}S_{T}(p_1,p_2,p_3)}
\eqn\ST{
S_{T}(p_1,p_2,p_3)={\G ( {1\ov 2} +2p_1p_2)\G ( {1\ov 2} +2p_1p_3 )
\G ( {1\ov 2} +2p_2p_3)\G (2- p_1^2 - p_2^2 - p_3^2  ) 
\ov \pi^{{3\ov 2}}\G (1-2p_1^2)\G (1-2p_2^2)\G (1-2p_3^2)}=
}
$$
=
{\G ( {1\ov 2} +p_1^2 - p_2^2 - p_3^2 )
\G ( {1\ov 2} +p_2^2 - p_1^2 - p_3^2 )
\G ( {1\ov 2} +p_3^2 - p_2^2 - p_1^2 )
\G (2 - p_1^2 - p_2^2 - p_3^2 ) 
\ov \pi^{{3\ov 2}}\G (1-2p_1^2)\G (1-2p_2^2)\G (1-2p_3^2)},
$$
where we used the momentum conservation $p_1 + p_2 + p_3 = 0.$
The tachyon cubic term was also computed in \km, and expressed in 
terms of the hypergeometric function ${}_3F_2$. We couldn't show 
that the expression \TTTs\ we obtained coincides with the one in \km. 
The cubic term exhibits a rather unusual momentum dependence. 
One can easily see that if one or two tachyons are on shell, 
$p_i^2= {1\ov 2}$,
then the tachyon cubic term vanishes. This implies that any exchange
diagram involving the cubic  term vanishes, and, in particular,
the 4-point tachyon scattering amplitude should be given just by
the corresponding quartic term in the effective action. This is 
consistent with the fact  that the on-shell effective action coincides
with the sigma model partition function,
that is a generating functional
of perturbative tachyon amplitudes. On the other hand
if all the three tachyons are on shell, i.e. we are computing the 
3-point tachyon amplitude, we have an ill-defined expression
$0 \times \infty $. This ambiguity is obviously a manifestation
of the Moebius volume infinity \rf{\ts,\pol}.  One may
regularize the cubic term by adding to the quadratic action a soft mass
term $m^2T^2$. Then it is easy
to see that the 3-point amplitude has a finite limit at $m^2\to 0$.
We expect that
any n-point term in the tachyon effective action has
the same property, and that explains the absence of the contribution
of exchange diagrams to perturbative tachyon amplitudes.
\newsec{Massive string modes}
In this section we proceed with the study of the boundary sigma
model by turning on the vector field and the massive
symmetric tensor field of the second rank. We will compute
the quadratic action for tachyon, vector
and massive symmetric tensor fields, and show that it exhibits
the required mass-shell conditions for all the fields.
Although the inclusion of the massive tensor field leads to
nonrenormalizability of the sigma model, the partition function and
the action can be made finite by adding proper boundary terms.

The boundary sigma model with tachyon, vector and symmetric tensor
turned on is described by the boundary interaction
\eqn\tab{
S_{\pa \Sigma } =  \int_0^{2\pi} \, {d\tau
\over 2\pi}\biggl( {1\ov\ee}T(x(\ta )) - {i\over 2} A_{\m} (x(\ta
))\dot{x}^\m + \ee B_{\m\n}\dot{x}^\m\dot{x}^\n\,\biggr) .
}
Expanding the partition function \zta\ up to
the second order in the fields, we get
\eqn\ztabc{
Z(T,A)= \la 1\ra + T + B + TT + AA + TB + BB,
}
where  $T$ and $TT$ are given by \T\ and \TT, and
\eqn\B{
B=-\ee \int \, {d\tau\ov 2\pi}\
\la\ B_{\m\n}\dot{x}^\m\dot{x}^\n \ \ra  =
- {1\ov\ee}\int\ d^dx\ B_\m^\m (x)
}
\eqn\AA{
AA =   -{1\ov 8}
\int \, {d\tau\ov 2\pi}{d\ta ' \ov 2\pi}\
\la\ A_\m (x(\ta ))\dot{x}^\m(\ta )
A_{\m'} (x(\ta '))\dot{x}^{\m'}(\ta' )\
\ra ,
}
\eqn\TB{
TB =
\int \, {d\tau\ov 2\pi}{d\ta ' \ov 2\pi}\
\la\ T(x(\ta ))B_{\m\n} (x(\ta '))\dot{x}^\m(\ta' )\dot{x}^{\n}(\ta' )\
\ra ,
}
\eqn\BB{
BB = {1\ov 2}\ee^2
\int \, {d\tau\ov 2\pi}{d\ta ' \ov 2\pi}\
\la\ B_{\m\n}(x(\ta ))B_{\m'\n'} (x(\ta '))
\dot{x}^\m(\ta )\dot{x}^{\n}(\ta )
\dot{x}^{\m'}(\ta' )\dot{x}^{\n'}(\ta' )\
\ra ,
}
Comparing \T\ and \B, we see that we have to shift
$T$ to remove the term
linear in $B_\m^\m$
\eqn\dTtoB{
T(x)\to T(x)- B_\m^\m (x) .
}
Computation of the quadratic terms is done by using
formulas from Appendix B. The expression for the tachyon
quadratic term is given in Appendix A. For the vector field
quadratic term one gets
\eqn\AAf{
AA =
- {1\ov 8}(2\pi )^{d}\int d^dp\ F_{\m\n}(p)F_{\m\n}(-p) \left[
\ee^{4p^2} {\Gamma ( - {1\ov 2} - 2p^2 )\ov \G ( {1\ov 2} )
\G (1-2p^2)} +{1\ov\ee}{\Gamma ( {1\ov 2} + 2p^2 )\ov \G ( {1\ov 2} )
\G (2+2p^2)}  \right],
}
The first term in \AAf\ coincides with the one computed in \ts\ by
using the analytical continuation in momenta.
We also see that \AAf\ contains a power divergent term which can be
removed by the tachyon redefinition
\eqn\dTtoFF{
 T(x)\to T(x) -{1\ov 8 (2\pi)^d}\int d^dy\ d^dp\ F_{\m\n}(x+{y\ov 2})
 F_{\m\n}(x-{y\ov 2}) \e^{ip\cdot y } {\Gamma ( {1\ov 2} + 2p^2 )\ov
 \G ( {1\ov 2} )\G (2+2p^2)}
 }
This tachyon redefinition is a generalization to all orders in
derivatives of the one discussed in \tsr. After the redefinition
the renormalization of the quadratic term in the partition function
is done by renormalizing the
bare vector field $A$ as
\eqn\vren{
A(p)\ \ee^{2p^2} = A_{R}(p)\ \ee_{{}_R}^{2p^2} ,}
that gives the vector field beta-function
\eqn\vecb{
\b_A (p) = 2p^2 A(p).
}
Then by using \sz\ and the beta function,
one finds the vector field
quadratic action\foot{Here and in what follows
we write quadratic actions for the renormalized fields, and
omit the subscript $R$ on $T$ in all the formulas.}
\eqn\AAs{
S_{AA}
={1\ov 4}(2\pi )^{d}\int d^dp\ F_{\m\n}(p)F_{\m\n}(-p)
\ee_{{}_R}^{4p^2} {\Gamma ( {1\ov 2} - 2p^2 )\ov
\G ( {1\ov 2} )\G (1-2p^2)} .}
Expanding the quadratic term in powers of $p^2$, we obtain
the usual ${1\ov 4} F^2$ term with the conventional coefficient.

The cross term $TB$ is given by
\eqn\TBf{
TB = - 2(2\pi )^{d}\int d^dp\ T(p)p^\m p^\n B_{\m\n}(-p) \left[
\ee^{4p^2} {\Gamma ( - {1\ov 2} - 2p^2 )\ov \G ( {1\ov 2} )
\G (1-2p^2)} +{1\ov\ee}{\Gamma ( {1\ov 2} + 2p^2 )\ov
\G ( {1\ov 2} )\G (2+2p^2)}  \right]
}
$$
+ {1\ov  \ee^2}(2\pi )^{d}\int d^dp\ T(p)B_\m^\m(-p)\ \ee^{4p^2}
{\Gamma (
{1\ov 2} - 2p^2 )\ov \G ( {1\ov 2} ) \G (1-2p^2)}   .
$$
It can be easily checked that the last term is canceled by
a similar term coming from $TT$ after the tachyon shift \dTtoB.
The power divergent term is removed by a redefinition  similar to
\dTtoFF
$$
T(x)\to T(x) +{2\ov  (2\pi)^d}\int d^dy\ d^dp\ T(x+{y\ov 2})
\pa^\m\pa^\n B_{\m\n}(x-{y\ov 2}) \e^{ip\cdot y }
{\Gamma ( {1\ov 2} + 2p^2 )\ov  \G ( {1\ov 2} )\G (2+2p^2)}
$$
The renormalization of the first term in \TBf\ is done
by renormalizing the bare  $B_{\m\n}$ as
\eqn\bren{
 B_{\m\n} (p)\ \ee^{2p^2+1} =
B^{R}_{\m\n}(p)\ \ee_{{}_R}^{2p^2},}
that gives the beta-function for $B_{\m\n}$
\eqn\bb{
\b_B (p) = (2p^2+1) B(p).
}
The corresponding term in the action is then found by using
\sz, the tachyon beta-function \tacb\ and \bb
\eqn\TBs{
S_{TB} = 4(2\pi )^{d}\int d^dp\ T(p)p^\m p^\n B_{\m\n}(-p)
\ee_{{}_R}^{4p^2} {\Gamma (  {1\ov 2} - 2p^2 )\ov \G ( {1\ov 2} )
\G (1-2p^2)}
}
The term quadratic in $B_{\m\n}$ has the form
\eqn\BBf{
BB =
{1\ov 2}(2\pi )^{d}\int d^dp\ \ee^{4p^2+2}\Biggl[
{1\ov \ee^4}B_\m^\m(p)B_{\n}^{\n}(-p)
{\Gamma ( {1\ov 2} - 2p^2 )\ov \G ( {1\ov
2} ) \G (1-2p^2)} }
$$
- {1\ov \ee^2}p^\m p^\n B_{\m\n} (p)B^{\rho}_{\rho}(-p)
{4\ \Gamma ( -{1\ov 2} - 2p^2 )\ov \G ( {1\ov 2} )
\G (1-2p^2)}
+ B_{\m\n} (p)B^{\m\n}(-p) {2\
\Gamma ( -{3\ov 2} - 2p^2 )\ov
\G ( {1\ov 2} ) \G (-1-2p^2)}
$$
$$ + p^\m p^\n B_{\m\n} (p)B^{\rho}_{\rho}(-p) {8\
(1+p^2)\Gamma ( -{3\ov 2} - 2p^2 )\ov \G ( {1\ov 2} )
\G (-2p^2)}
$$
$$
+ {12\ \Gamma ( -{3\ov 2} - 2p^2 )\ov \G ( {1\ov 2} )
\G (1-2p^2)}\left( p^\m p^\n B_{\m\n} (p) p^\rho p^\s B_{\rho\s} (-p) -
{4\ov 3}p^2 p^\n p^\s B_{\m\n} (p) B_{\m\s} (-p) \right)
$$
$$
+\ee^{-4p^2-3} B_{\m\n} (p) B_{\rho\s} (-p)
\tilde{K}^{\m\n\rho\s}(p)
+\ee^{-4p^2-1}B_{\m\n}(p) B_{\rho\s} (-p)K^{\m\n\rho\s}(p)
\Biggr]. $$
Here $K^{\m\n\rho\s}(p)$ and  $\tilde{K}^{\m\n\rho\s}(p)$
are tensors which do not depend on $\ee$.
Explicit expressions for the last two terms are given in Appendix B.

It is not difficult to check that the first two terms in \BBf\ are
canceled by the corresponding terms coming from $TT$ and $TB$ after
the tachyon shift \dTtoB.  The next three terms are finite in
terms of the renormalized fields $B^R_{\m\n}$, eq. \bren.
The power divergent term (with $\tilde{K}$)
is again removed by a tachyon
shift of the form \dTtoFF. The last term reflects the
nonrenormalizability  of the model. Being expressed in terms of
the renormalized fields, it takes the form
\eqn\Bdiv{
{1\ov 2}(2\pi )^{d}\int d^dp\
{1\ov \ee} \left({\ee\ov\ee_{{}_R}}\right)^{-4p^2}
B^{R}_{\m\n}(p) B^R_{\rho\s}(-p)K^{\m\n\rho\s}(p) ,
}
which shows explicitly that it diverges as $\ee\to 0$. Thus,
to have a finite partition function one has to
cancel the term. It can be done by adding to the boundary
interaction \tab\ some higher-derivative term. The requirement
that the additional boundary term has a minimal number of
derivatives fixes the form  of the term to be
\eqn\adbt{
 S'_{\pa \Sigma } =  \int_0^{2\pi} \, {d\tau
\over 2\pi}\ \ee^3 K(B(x(\ta ))) \left(
\dot{x}^\m\dot{x}_\m \dot{x}^\n\dot{x}_\n -{2(d+2)\ov 3}
\ddot{x}^\m\ddot{x}_\m
\right)
}
where $K(B(x(\ta )))$ is the following functional of bare $B_{\m\n}$
\eqn\kb{
 K(B(x(\ta )))=-{3\ov 2d(d+2)}{1\ov 2 (2\pi)^d}\int d^dy\ d^dp\
B_{\m\n}(x(\ta )+{y\ov 2}) B_{\rho\s}(x(\ta )-{y\ov 2})
\e^{ip\cdot y}K^{\m\n\rho\s}(p),
}
and the constant ${2(d+2)\ov 3}$ in \adbt\
ensures the absence
of terms diverging as ${1\ov \ee}$.
Thus, despite the nonrenormalizability of the model,
there is a well-defined and consistent way of removing
divergent terms.

Now, having the finite partition function, and using
the tensor field beta-function \bb, we find the
quadratic action for tachyon and $B_{\m\n}$
\eqn\STB{
S = (2\pi )^{d}\int d^dp\ \ee_{{}_R}^{4p^2}\Biggl[
 - T(p)T(-p) {\Gamma ( {3\ov 2} -
2p^2 )\ov \G ( {1\ov 2} ) \G (1-2p^2)} +
 4 T(p)p^\m p^\n B_{\m\n}(-p)
 {\Gamma (  {1\ov 2} - 2p^2 )\ov
\G ( {1\ov 2} ) \G (1-2p^2)}}
$$
-2 B_{\m\n} (p)B^{\m\n}(-p) {\Gamma ( -{1\ov 2} - 2p^2 )\ov
\G ( {1\ov 2} )
\G (-1-2p^2)}-
 8 p^\m p^\n B_{\m\n} (p)B^{\rho}_{\rho}(-p){(1+p^2)
 \Gamma ( -{1\ov 2} - 2p^2
)  \ov \G ( {1\ov 2} )
\G (-2p^2)}
$$
$$
- 12 {\Gamma ( -{1\ov 2} - 2p^2 )\ov \G ( {1\ov 2} )
\G (1-2p^2)}\left( p^\m p^\n B_{\m\n} (p) p^\rho p^\s B_{\rho\s} (-p) -
{4\ov 3}p^2 p^\n p^\s B_{\m\n} (p) B_{\m\s} (-p) \right)  \Biggr].
$$
Let us recall that all the fields in \STB\ are renormalized.
We can remove the cross term $TB$ by shifting $T$
$$
T(p)\to T(p)+ {4\ov 1-4p^2}p^\m p^\n B_{\m\n}(p)
$$
Then we have a quadratic action for $B_{\m\n}$ which can be easily
shown to lead to the usual on-shell conditions
$$
2p^2 = -1, \ \ \ \ B_\m^\m = 0,\ \ \ \  p^\n B_{\m\n}(p)=0.
$$
The mass shell condition $2p^2 = -1$ immediately follows from
the term
$$
 B_{\m\n} (p)B^{\m\n}(-p) {\Gamma ( -{1\ov 2} - 2p^2 )\ov
 \G ( {1\ov 2} )
\G (-1-2p^2)}
$$
in the quadratic action.

Some comments are in order. It seems that one can make
any higher-order term in the sigma model partition function finite
by using the same procedure: one first removes power divergent terms
by a tachyon redefinition, then one
assumes that at any order in the fields their renormalization
is given by the formulas \tren, \vren\ and \bren, and expresses
the partition function in terms of the renormalized fields, and,
finally, one cancels all remaining singular terms by adding proper
higher-derivative boundary terms. Moreover, it seems possible to turn on
all massive modes up to some mass level $k$, and to perform all
computations in the same way as was done for $B_{\m\n}$.
It will be necessary to shift lower level massive modes to remove
power singularities coming from more massive modes, and to assume
that a massive mode $B_{(k)}$ of level $k$ is renormalized as
\eqn\mren{
B_{(k)} (p)\ \ee^{2p^2+k} =
B^{R}_{(k)}(p)\ \ee_{{}_R}^{2p^2}.}
Although it is unclear if the level truncated effective
action constructed this way is unique up to fields redefinitions,
we should note that the quadratic action \STB\ was derived
unambiguously.

\newsec{Fermionic sigma model}
In this section we first compute the quadratic term in the
effective action for tachyon on an unstable D9-brane in type
II string theory, and show that it reproduces the correct
tachyon mass. Then, we turn on a massive symmetric tensor $B_{\m\n}$,
and show that there is a term linear in $B_\m^\m$, which remains finite
after renormalization.
\subsec{ Tachyon quadratic action}
Tachyon on an unstable
D9-brane in type II string theory is described
by sigma model with the action \rf{\ed,\kmm}
$$
S = S_{\Sigma } + S_{\pa \Sigma },
$$
where
$$
S_{\Sigma } = {1\over 4\pi}
\int \, d^2\, z \biggl( \pa x^\m \bar{\pa} x_\m
+ \psi^\m \bar{\pa}\psi_\mu + \tilde{\psi}^\mu\pa
\tilde{\psi}_\mu \biggr)
$$
and
$$
S_{\pa \Sigma }={1\ov 4}\int \, {d\tau\ov 2\pi} \,\biggl(
{1\ov\ee}T^2+{1\ov\ee}\left( \partial_\m T\psi^\m \right)
\partial_\tau^{-1}\left( \partial_\n T\psi^\n \right)
\biggr) .
$$
The tachyon effective action just coincides with the sigma model
partition function
\eqn\zt{
S(T)=Z(T)=\la {\rm e}^{-S_{\pa \Sigma }}\ra ,
}
and, therefore, its quadratic part is given by
\eqn\quad{
S_{TT}=\la  -{1\ov 4\ee}\int \, {d\tau\ov 2\pi} \,\biggl(
T(x(\ta ))^2+\left( \partial_\m T(x(\ta ))\psi^\m \right)
\partial_\tau^{-1}\left( \partial_\n T(x(\ta ))\psi^\n \right)
\biggr)\ra }
$$=  -{1\ov 4\ee}\int d^dx\ T^2 -
{1\ov 8\ee}\int \, {d\tau\ov 2\pi}d\ta ' \
\la\ \partial_\m T(x(\ta ))\psi^\m (\ta )\ev (\ta -\ta ')
\partial_\n T(x(\ta '))\psi^\n (\ta ')\ \ra ,
$$
where $\ev (\ta ) = +1$ for $ \ta > 0$ and $\ev (\ta ) =-1$
for $\ta < 0$.

To compute the quadratic term one needs the boundary fermionic
Green function
\eqn\fgr{
G_F^{\m\n}(\tau ):= \la \psi^\m(\tau)\psi^\nu(0)\ra
=\d^{\m\n}G_F(\tau )
=-\d^{\m\n}{2 \sin\left({\ta\ov 2}\right)\ov
\sin^2\left({\ta\ov 2}\right)+\ee^2} ,
}
and the bosonic Green function \bgrff.
By using the Green functions and
the momentum representation \Tp, one finds
\eqn\tt{
S_{TT}^{(1)}=- \int d^dx {1\ov 4\ee}\int \, {d\tau\ov 2\pi}\
\la\ T(x(\ta ))T(x(\ta ))\ra =
-{1\ov 4\ee}(2\pi )^{d}\int d^dp T(p)T(-p)
}
and
\eqn\dtdt{
S_{TT}^{(2)}=- \int d^dx {1\ov 8\ee}\int \, {d\tau\ov 2\pi}d\ta ' \
\la\ \partial_\m T(x(\ta ))\psi^\m (\ta )\ev (\ta -\ta ')
\partial_\n T(x(\ta '))\psi^\n (\ta ')\ \ra }
$$={1\ov 4}(2\pi )^{d+1}\int d^dp T(p)T(-p)\ p^2\ \ee^{4p^2-1}\
\int_0^{2\pi}{d\tau\ov 2\pi}\sin\bigl( {\ta \ov 2} \bigr)
\biggl[\sin^2\bigl( {\ta \ov 2} \bigr)+ \ee^2\biggr]^{
- 2 p^2 - 1 } .
$$
Computing the integral over $\ta$, we obtain in the limit $\ee\to 0$
\eqn\dtdtf{
S_{TT}^{(2)}=
{1\ov 4}(2\pi )^{d}\int d^dp\ T(p)T(-p)\left( {1\ov\ee}
- \ee^{4p^2-1}
{\G ( {1\ov 2} )\Gamma (  1-2p^2  )\ov
\G ( {1\ov 2} - 2p^2 )}\right) .}
Combining \tt\ and \dtdtf, we derive the quadratic
term in the partition
function
\eqn\pff{
S_{TT}=Z_{TT}^{(1)}+Z_{TT}^{(2)}
=-{1\ov 4}(2\pi )^{d}\int d^dp\ T(p)T(-p)\
 \ee^{4p^2-1}
{\G ( {1\ov 2} )\Gamma (  1-2p^2  )\ov
\G ( {1\ov 2} - 2p^2 )} .}
Thus, we see that the power divergent terms cancel each other,
and the quadratic term can be made finite by the
tachyon renormalization
\eqn\tfren{
T (p)\ \ee^{2p^2-{1\ov 2}} =
T_{R}(p)\ \ee_{{}_R}^{2p^2}.}
Then,
the integrand exhibits zero
at $p^2 = {1 \ov 4}$, which is the correct tachyon mass in open
superstring theory. We also see that if one would use
the definition \sz\ for  the tachyon action and the tachyon
beta-function, $\b_T(p)=2p^2-{1\ov 2}$,
one would get the term
\eqn\sha{
\tilde{S}_{TT}= -(2\pi )^{d}\int d^dp\ T(p)T(-p)\
 \ee^{4p^2-1}p^2
{\G ( {1\ov 2} )\Gamma (  1-2p^2  )\ov
\G ( {1\ov 2} - 2p^2 )}  ,}
which exhibits an additional zero at $p^2 = 0$.
Expanding the quadratic action  in powers of $p$ we get
$$S_{TT}=
{1\ov 4}(2\pi )^{d}\int d^dp T(p)T(-p)\ \left(
-1 +(8\log 2 - 4\log 2\ee_{{}_R} ) p^2 \right) $$
If we choose $\ee_{{}_R} ={1\ov 2}$ in this expression,
we reproduce the quadratic
term that follows from the action found in \kmm.
If we choose $\ee_{{}_R} ={2\ov \e}$ we reproduce the 2-derivative action
with the correct tachyon mass.

\subsec{Massive symmetric tensor}
In this subsection we set tachyon $T=0$, and only consider
a massive symmetric tensor field $B_{\m\n}$.
The boundary interaction describing the field is given by
\eqn\bib{
S_{\pa \Sigma }={1\ov 4}
\int \ {d\tau\ov 2\pi} \ d\theta \
 \ee B_{\m\n}(X) D^2X^\m DX^\n  \ .
}
Here
$$X^\m = x^\m(\ta )+\theta\psi^\m (\ta ) ,\ \ \ \ D=\partial_\theta
+\theta\partial_\tau
$$
are matter superfields and a supercovariant derivative,
respectively. Integrating over $\theta$, we get the component form
of the boundary term
\eqn\bibc{
S_{\pa \Sigma }={1\ov 4}
\int \, {d\tau\ov 2\pi} \ \biggl(
  \ee B_{\m\n} \left( \dx^\m \dx^\n +
\dot{\psi}^\m \psi^\n \right) +
\ee \pa_\r B_{\m\n}\dx^\m \psi^\r \psi^\n \ \biggr) .
}
To compute correlators in this model we need regularized
bosonic and fermionic boundary Green functions. We cannot use
the functions \bgrff\ and \fgr\ because they do not preserve
1-d supersymmetry. They violate supersymmetry only at order $o(\ee)$,
and by this reason we could use them to compute the tachyon quadratic
action, but once we turn on a nonrenormalizable interaction we
are to use Green functions exactly preserving supersymmetry.\foot{
Strictly speaking, the supersymmetry is spontaneously broken by
the antiperiodicity of the boundary fermions \an. What we mean
by saying that
the regularization is supersymmetric is that it
would preserve the supersymmetry if the fermions were periodic.}
A possible choice is
\eqn\bgr{
G_B^{\ee}(\tau)=
4\,\sum_{k=1}^\infty \ {\rm e}^{-2k\ee }{\cos k\ta\over k } =
-2\left( \log\left( \sin^2\left({\ta\ov 2}\right)+
\sinh^2\left(\ee\right)\right) +\log 4 -2\ee \right) ,
}
and
\eqn\fgrff{
G_F^{\ee}(\tau)
=-4\ \sum_{r=1/2}^\infty{\rm e}^{-2r\ee}\sin r\ta =
-{2 \cosh\left(\ee\right) \sin\left({\ta\ov 2}\right)\ov
\sin^2\left({\ta\ov 2}\right)+\sinh^2\left(\ee\right)}.}
By using the Green functions one can easily compute terms linear in
$B_\m^\m$ in the effective action \zt
\eqn\ltb{
S^{(1)}(B) = {1\ov 8}\int d^{10}x\ \ee B_\m^\m .
}
Renormalization of $B_{\m\n}$ (see, \bren ) just absorbs $\ee$, and
one gets a finite linear term. Contrary to the bosonic case,
this linear term cannot be removed by a tachyon redefinition because
there is no term linear in tachyon in the fermionic case.
On the other hand, this term would violate the usual on-shell
conditions for $B_{\m\n}$. Thus we have to cancel it somehow.
A possible way is to add a proper higher-derivative term to
the disc boundary. It is not difficult to show that the simplest
choice is
\eqn\adbt{
S'_{\pa \Sigma }={1\ov 4}
\int \, {d\tau\ov 2\pi} \, d\theta
\bigl( -{1\ov 120} \ee^3
B_{\m}^{\m}(X)D^2X^\n D^2X_\n D^2X^\rho DX_\rho
 \bigr)}
$$
 =
-{\ee^3\ov 480}
\int \, {d\tau\ov 2\pi} \
\biggl( B_{\m}^{\m}\left(
\dx^\n \dx_\n \dx^\rho \dx_\rho +
\dx^\n \dx_\n\dot{\psi}^\rho \psi_\rho
+ 2\dx^\n \dx_\rho \dot{\psi}^\n \psi^\rho
\right)+ \pa_\s B_{\m}^{\m}
\dx^\n \dx_\n \dx^\rho\psi^\s \psi_\rho
 \biggr)   .
$$
The coefficient in front of the term was fixed by using
$$\la \dx^\n \dx_\n \dx^\rho \dx_\rho +
\dx^\n \dx_\n\dot{\psi}^\rho \psi^\rho
+ 2\dx^\n \dx_\rho \dot{\psi}^\n \psi^\rho  \ra =
30 ( -{2\ov\ee^2} +{7\ov 6} + o(\ee ) ) .$$
Although one can cancel the linear term
by adding to the disc boundary such a higher-derivative term,
the procedure does not look completely satisfactory. The choice
of a higher-derivative term is not unique, and different choices
would lead to different off-shell actions for $B_{\m\n}$, and
it is unclear how to argue that they are equivalent.

\newsec{ Acknowledgements}
The author is grateful to A. Tseytlin  for numerous valuable
discussions and comments on the manuscript.
The work was supported by the U.S. Department of Energy under grant
No. DE-FG02-91ER-40690 and in part by RFBI grant N99-01-00190.

\appendix{A}{Tachyon cubic term}
Here we list the results obtained for
the quadratic and cubic terms in
the partition function
\eqn\TTr{
TT=  {1\ov 2\ee^2}(2\pi )^{d}\int d^dp\ T(p)T(-p)\ \ee^{4p^2}
{\Gamma ( {1\ov 2}- 2p^2 )\ov \G ( {1\ov 2} ) \G (1-2p^2)}  ,
}
\eqn\TTTr{
TTT= - {(2\pi )^d\ov 6\ee^3}\int d^dp_i \d (\sum_i p_i)
T(p_1 ) T(p_2)T(p_3)\
\left( \ee^2\right)^{\sum_i p_i^2}K_{T}(p_1,p_2,p_3)}
\eqn\KT{
K_{T}(p_1,p_2,p_3)={\G ( {1\ov 2} +2p_1p_2)\G ( {1\ov 2} +2p_1p_3 )
\G ( {1\ov 2} +2p_2p_3)\G (1- p_1^2 - p_2^2 - p_3^2  ) 
\ov \pi^{{3\ov 2}}\G (1-2p_1^2)\G (1-2p_2^2)\G (1-2p_3^2)}
}
$$
=
{\G ( {1\ov 2} +p_1^2 - p_2^2 - p_3^2 )
\G ( {1\ov 2} +p_2^2 - p_1^2 - p_3^2 )
\G ( {1\ov 2} +p_3^2 - p_2^2 - p_1^2 )
\G (1 - p_1^2 - p_2^2 - p_3^2 ) 
\ov \pi^{{3\ov 2}}\G (1-2p_1^2)\G (1-2p_2^2)\G (1-2p_3^2)},
$$
The derivation of the quadratic terms is straightforward, and will be 
omitted here.

To compute the cubic term $TTT$ we first note, by using the momentum 
representation \Tp\ and the bosonic Green
function \bgrff, that the kernel
$K_T$ is given by
\eqn\KTt{
K_{T}(p_1,p_2,p_3)=\left( \ee^2\right)^{-\sum_i p_i^2}
\la \e^{ip_1\xi(\ta_1)}\e^{ip_2\xi(\ta_2)}\e^{ip_3\xi(\ta_3)}\ra }
$$
=\int \, {d\tau_1\ov 2\pi}{d\ta_2 \ov 2\pi}{d\ta_3 \ov 2\pi}\
\biggl[\sin^2\bigl( {\ta_{12} \ov 2} \bigr)+
\ee^2\biggr]^{ 2 p_1p_2 }
\biggl[\sin^2\bigl( {\ta_{13} \ov 2} \bigr)+
\ee^2\biggr]^{ 2 p_1p_3 }
\biggl[\sin^2\bigl( {\ta_{23} \ov 2} \bigr)+
\ee^2\biggr]^{ 2 p_2p_3 }
$$
where $\ta_{ij}=\ta_i-\ta_j$.

We are interested in the limit $\ee\to 0$ of the integral.
Since the limit exists for
small values of $p_i$, all we have to do is to
 compute the integral for
$\ee =0$
$$
K_{T}(p_1,p_2,p_3)=\int \, {d\tau_1\ov 2\pi}{d\ta_2 \ov 2\pi}\
\biggl[\sin^2\bigl( {\ta_{12} \ov 2} \bigr)\biggr]^{ 2 p_1p_2 } 
\biggl[\sin^2\bigl( {\ta_{1} \ov 2} \bigr)\biggr]^{ 2 p_1p_3 } 
\biggl[\sin^2\bigl( {\ta_{2} \ov 2} \bigr)\biggr]^{ 2 p_2p_3 } .$$
To compute the integral we first make a change 
$$
\e^{i\ta_1}={x_1-i\ov x_1+i},\ \ \ \ \e^{i\ta_2}={x_2-i\ov x_2+i},$$
i.e. we transform the circles into straight lines. Then $K_T$
acquires the form
$$
K_T = {1\ov \pi^2}\int_{-\infty}^{\infty}
{dx_1 dx_2\over (1+x_1^2)(1+x_2^2)}
\left( {(x_1- x_2)^2\over (1+x_1^2)(1+x_2^2)}\right)^{2p_1p_2}
\left( {1\over 1+x_1^2}\right)^{2p_1p_3}
\left( {1\over 1+x_2^2}\right)^{2p_2p_3}
$$
$$
={1\ov \pi^2}\int_{-\infty}^{\infty}dx_1 dx_2
\left( (x_1- x_2)^2\right)^{2p_1p_2}
\left( 1+x_1^2\right)^{2p_1^2-1}
\left( 1+x_2^2\right)^{2p_2^2-1}$$
By using the formula
\eqn\zma{
z^{-a}={1\ov\G (a)}\int_0^\infty dt \ t^{a-1}\e^{-tz}
}
we rewrite $K_T$ in the form
$$
K_T={1\ov \pi^2 \G (1-2p_1^2)\G (1-2p_2^2)}
\int_0^\infty dt_1dt_2 t_1^{-2p_1^2} t_2^{-2p_2^2} $$
$$\times
\int_{-\infty}^{\infty}dx_1 dx_2
\left( (x_1- x_2)^2\right)^{2p_1p_2}
\e^{-t_1(1+x_1^2)-t_2(1+x_2^2)}$$
Integrating over $x_1,x_2$,  one gets
$$
K_T={\G ( {1\ov 2} +2p_1p_2)\ov \pi^{{3\ov 2}}
\G (1-2p_1^2)\G (1-2p_2^2)}
\int_0^\infty dt_1dt_2 t_1^{-2p_1^2-2p_1p_2- {1\ov 2} } 
t_2^{-2p_2^2-2p_1p_2 - {1\ov 2} } 
(t_1+t_2)^{2p_1p_2}
\e^{-t_1-t_2}$$
By using again \zma, we obtain
$$K_T={\G ( {1\ov 2} +2p_1p_2)\ov \pi^{{3\ov 2}}
\G (-2p_1p_2)\G (1-2p_1^2)
\G (1-2p_2^2)}
\int_0^\infty dx x^{-2p_1p_2-1}$$
$$\times dt_1dt_2 \e^{-x(t_1+t_2)} 
t_1^{-2p_1^2-2p_1p_2- {1\ov 2} } t_2^{-2p_2^2-2p_1p_2 - {1\ov 2} } 
\e^{-t_1-t_2}$$
Integrating over $t_1,t_2$, we finally arrive at
$$K_T={\G ( {1\ov 2} +2p_1p_2)\G ( {1\ov 2} -2p_1^2 -2p_1p_2)\
G ( {1\ov 2} -2p_2^2 -2p_1p_2) 
\ov \pi^{{3\ov 2}}\G (-2p_1p_2)\G (1-2p_1^2)\G (1-2p_2^2)}
\int_0^\infty dx {(1+x)^{2p_1^2+4p_1p_2+2p_2^2-1}\ov x^{2p_1p_2+1}} 
$$
$$={\G ( {1\ov 2} +2p_1p_2)\G ( {1\ov 2} -2p_1^2 -2p_1p_2) )
\G ( {1\ov 2} -2p_2^2 -2p_1p_2)
\G (1 +2p_1p_2-2p_3^2) 
\ov \pi^{{3\ov 2}}\G (1-2p_1^2)\G (1-2p_2^2)\G (1-2p_3^2)}$$
$$=
{\G ( {1\ov 2} +p_1^2 - p_2^2 - p_3^2 )
\G ( {1\ov 2} +p_2^2 - p_1^2 - p_3^2 )
\G ( {1\ov 2} +p_3^2 - p_2^2 - p_1^2 )
\G (1 - p_1^2 - p_2^2 - p_3^2 ) 
\ov \pi^{{3\ov 2}}\G (1-2p_1^2)\G (1-2p_2^2)\G (1-2p_3^2)}$$
where we used the momentum conservation
$$p_1 + p_2 + p_3 = 0.$$
The computation of $TTT$ was obviously equivalent to the one done by
the analytical continuation in momenta. One may ask if one can
use the analytical continuation to compute the cubic term $AAT$
involving the open string vector field. The result of the computation
is given by
\eqn\AATr{
AAT= -{(2\pi )^d\ov 2\ee}\int d^dp_i \d (\sum_i p_i)
A_{\m_1}(p_1)A_{\m_2}(p_2)T(p_3)\
\left( \ee^2\right)^{\sum_i p_i^2}K_{A}(p_1,p_2,p_3) }
$$
\times
\left( p_1^2p_2^2\left(\d^{\m_1\m_2}-4 p_2^{\m_1}p_1^{\m_2}\right)+
\left( 1-4p_1p_2\right)\left( p_1p_2p_1^{\m_1}p_2^{\m_2}-
p_2^2p_1^{\m_1}p_1^{\m_2}-p_1^2p_2^{\m_1}p_2^{\m_2}\right)\right)
$$
$$=-{(2\pi )^d\ov 2\ee}\int d^dp_i \d (\sum_i p_i)
T(p_3)\ \left( {\ee^2\ov 4}\right)^{\sum_i p_i^2}K_{A}(p_1,p_2,p_3)
\times
$$
$$
\left( {1\ov 2} p_1p_2 F_{\m\n}(p_1)F^{\m\n}(p_2) -
F_{\m\n}(p_1)F^{\m\rho}(p_2) \left( p_1^\n p_2^\rho +p_2^\n p_1^\rho
\right) -4 F_{\m\n}(p_1)F_{\rho\s}(p_2)
p_1^{\m}p_2^{\n}p_1^{\rho}p_2^{\s}\right) ,
$$
\eqn\KA{
K_{A}(p_1,p_2,p_3)={\G ( {1\ov 2} +2p_1p_3)\G ( {1\ov 2} +2p_2p_3)
\G ( - {1\ov 2} +2p_1p_2)\G (- p_1^2 - p_2^2 - p_3^2  )
\ov \pi^{{3\ov 2}}\G (1-2p_1^2)\G (1-2p_2^2)\G (1-2p_3^2)}
}
$$=
{\G ( {1\ov 2} +p_1^2 - p_2^2 - p_3^2 )
\G ( {1\ov 2} +p_2^2 - p_1^2 - p_3^2 )
\G ( - {1\ov 2} +p_3^2 - p_2^2 - p_1^2 )
\G ( - p_1^2 - p_2^2 - p_3^2 )
\ov \pi^{{3\ov 2}}\G (1-2p_1^2)\G (1-2p_2^2)\G (1-2p_3^2)},
$$
We see that the kernel $K_A$ has a pole at $p_i=0$, and, therefore,
does not admit the expansion in powers of momenta.
That probably means that one cannot use the analytical continuation to
compute the cubic term $AAT$. Note, however, that the term $S_{AAT}$
in the effective action \sz\ does admit
the expansion in powers of momenta.

\appendix{B}{Useful formulas}
In this section $G(\ta )\equiv G_B(\ta )\equiv G_\ta ,\ \
G(0)\equiv G_0,
\ \ {d\ov d\ta}G(\ta )=G'_\ta,\ \  {d^2\ov d\ta^2}G(\ta ) =G''_\ta  $.
\eqn\EE{
\int \, {d\tau\ov 2\pi} {d\tau'\ov 2\pi}\la
\e^{ip\cdot (\xi (\ta )- \xi (\ta' ))} \ra =
\int \, {d\tau\ov 2\pi} \e^{p^2 (G_\ta - G_0 )}}
$$
=\int \, {d\tau\ov 2\pi}\ \ee^{4p^2 }
\biggl[\sin^2\bigl( {\ta \ov 2} \bigr)+ \ee^2\biggr]^{-2p^2}  $$
$$
={1\ov\pi} \int_0^{1}{dy\ov \sqrt{y(1 - y)}}
 {1\ov (1+\ee^{-2}y)^{2p^2}} =
F \left( {1\ov 2},2p^2;1;-{1\ov\ee^2} \right)
$$
$$
= \ee^{4p^2 }
{\Gamma ( {1\ov 2} - 2p^2 )\ov \G ( {1\ov 2} )
\G (1-2p^2)}F\left(2p^2, 2p^2;{1\ov 2}+2p^2;-\ee^2\right)$$
$$
\ \ \ \ \ \ + \
 \ee {\Gamma ( -{1\ov 2} + 2p^2 )\ov \G ( {1\ov 2} )
\G (2p^2)}F\left({1\ov 2},{1\ov 2};{3\ov 2}-2p^2;-\ee^2\right) .
$$
Here, and in what follows, we use the formula
$$
 F(a, b; c; z) =
  {\Gamma(c)\Gamma(b - a)\ov\Gamma(b)\Gamma(c - a)}{1\ov (-z)^a}
F(a,1 - c + a; 1 - b + a; {1\ov z})
$$
$$\ \ \ \ \ \ \ \ \ + \
    {\Gamma(c)\Gamma(a - b)\ov \Gamma(a)\Gamma(c - b)}{1\ov (-z)^b}
F(b,1 - c + b; 1 - a + b; {1\ov z}) .
$$
\eqn\EExx{
\int \, {d\tau\ov 2\pi} {d\tau'\ov 2\pi}\la
\e^{ip\cdot (\xi (\ta )- \xi (\ta' ))}
\dot{\xi}^\m (\ta ) \dot{\xi}^\n (\ta' )\ra =
\left( \d^{\m\n} p^2 - p^\m p^\n \right)
\int \, {d\tau\ov 2\pi} \e^{p^2 (G_\ta - G_0 )}
\left ({d\ov d\ta}G(\ta )\right)^2}
$$
\int \, {d\tau\ov 2\pi} \e^{p^2 (G_\ta - G_0 )}
\left ({d\ov d\ta}G(\ta )\right)^2 =
$$
$$
\int \,
{d\tau\ov 2\pi} \ee^{4p^2 }
\biggl[\sin^2\bigl( {\ta \ov 2} \bigr)+ \ee^2\biggr]^{-2p^2-2}
\left (-2\sin\bigl( {\ta \ov 2} \bigr)
\cos\bigl( {\ta \ov 2} \bigr)\right)^2
$$
$$
=
{4\ov\pi} {1\ov\ee^4} \int_0^{1}dy{\sqrt{y(1 - y)}\ov
(1+\ee^{-2}y)^{2p^2+2}} ={1\ov 2} {1\ov\ee^4}
F \left( {3\ov 2},2+2p^2;3;-{1\ov\ee^2} \right)
$$
$$
=
\ee^{4p^2 }
{2\ \Gamma ( -{1\ov 2}-2p^2)\ov\G ( {1\ov 2} )
\G (1-2p^2)}F\left(2+2p^2, 2p^2;{3\ov 2}+2p^2;-\ee^2\right)
$$
$$
\ \ \ \ \ \ + \
 {1\ov\ee}{2\Gamma ( {1\ov 2} + 2p^2 )\ov \G ( {1\ov 2} )
\G (2+2p^2)}
F\left({3\ov 2},-{1\ov 2};{1\ov 2}-2p^2;-\ee^2\right)
$$
$$
=
\ee^{4p^2 }
{2\ \Gamma ( -{1\ov 2}-2p^2)\ov\G ( {1\ov 2} )
\G (1-2p^2)}-
\ee^{4p^2 +2 }
{4\ (1+p^2)\Gamma ( -{3\ov 2}-2p^2)\ov\G ( {1\ov 2} )
\G (-2p^2)}
$$
$$
\ \ \ \ \ \ + \
 {1\ov\ee}{2\ \Gamma ( {1\ov 2} + 2p^2 )\ov \G ( {1\ov 2} )
\G (2+2p^2)} -\ee {3\ \Gamma ( -{1\ov 2}+2p^2 )\ov 2 \G ({1\ov 2})
\G (2+2p^2)} .
$$

\eqn\EExxxx{
\int \, {d\tau\ov 2\pi} {d\tau'\ov 2\pi}\la
\e^{ip\cdot (\xi (\ta )- \xi (\ta' ))}
\dot{\xi}^\m (\ta ) \dot{\xi}^\n (\ta )
 \dot{\xi}^\rho (\ta' ) \dot{\xi}^\s (\ta' )\ra =
\int \, {d\tau\ov 2\pi} \e^{p^2 (G_\ta - G_0 )}\times}
$$
\Biggl[\d^{\m\n}\d^{\r\s} \left( G''_0\right)^2 +
 \left( \d^{\m\r}\d^{\n\s} + \d^{\m\s}\d^{\n\r}\right)
 \left( G''_\ta\right)^2
$$
$$
+
\left( \d^{\m\n}p^\r p^\s + \d^{\r\s}p^\m p^\n\right)
 G''_0\left( G'_\ta\right)^2
 + p^\m p^\n p^\r p^\s \left( G'_\ta\right)^4
 $$
 $$
 -{1\ov 3}p^2\left(\d^{\m\r}p^\n p^\s + \d^{\m\s}p^\n p^\r +
 \d^{\n\r}p^\m p^\s +
\d^{\n\s}p^\m p^\r\right)  \left( G'_\ta\right)^4  \Biggr] .
$$

\eqn\EEddGs{
 \int \, {d\tau\ov 2\pi} \e^{p^2 (G_\ta - G_0 )}
 \left( G''_\ta\right)^2 =}
 $$
  \int \,
{d\tau\ov 2\pi} \ee^{4p^2 }
\biggl[\sin^2\bigl( {\ta \ov 2} \bigr)+ \ee^2\biggr]^{-2p^2}
\left ({\sin^2\bigl( {\ta \ov 2}\bigr)\left( 1+ 2\ee^2\right)
- \ee^2 \ov \sin^2\bigl( {\ta \ov 2} \bigr)+
 \ee^2}\right)^2
$$
$$
={1\ov\pi}{1\ov\ee^8} \int_0^{1}{dy\ov \sqrt{y(1 - y)}}
 {\left( y \left( 1+ 2\ee^2\right)- \ee^2\right)^2 \ov
 (1+\ee^{-2}y)^{2p^2+4}}
 $$
 $$
 ={1\ov 32} {1\ov\ee^8} \Biggl[
2^5\ee^4 F \left( {1\ov 2},4+2p^2;1;-{1\ov\ee^2} \right)
$$
$$
-(2+4\ee^2)\left( 16\ee^2
F \left( {3\ov 2},4+2p^2;2;-{1\ov\ee^2} \right)-
 3(2+4\ee^2) F \left( {5\ov 2},4+2p^2;3;-{1\ov\ee^2} \right)
 \right)  \Biggr]
$$
$$
= \ee^{4p^2 }
{\Gamma ( -{3\ov 2}-2p^2)\ov\G ( {1\ov 2} )
\G (-1-2p^2)}+
 {1\ov\ee^3}{(3+6p^2+4p^4)\Gamma ({3\ov 2} + 2p^2)\ov\G ({1\ov 2} )
\G (4+2p^2)}-{1\ov\ee}{4p^2 \Gamma ({3\ov 2}+2p^2 )\ov \G ({1\ov 2})
\G (4+2p^2)}.
$$

\eqn\EEdGf{
 \int \, {d\tau\ov 2\pi} \e^{p^2 (G_\ta - G_0 )}
 \left( G'_\ta\right)^4 =}
 $$
  \int \,
{d\tau\ov 2\pi} \ee^{4p^2 }
\biggl[\sin^2\bigl( {\ta \ov 2} \bigr)+ \ee^2\biggr]^{-2p^2}
\left (-{2\sin\bigl( {\ta \ov 2}\bigr)\cos\bigl( {\ta \ov 2}\bigr)
\ov
\sin^2\bigl( {\ta \ov 2} \bigr)+  \ee^2}\right)^4
$$
$$
={16\ov\pi}{1\ov\ee^8} \int_0^{1}dy
{\left(y(1 - y)\right)^{{3\ov 2}}\ov
 (1+\ee^{-2}y)^{2p^2+4}}
 ={3\ov 8} {1\ov\ee^8}
  F \left( {5\ov 2},4+2p^2;5;-{1\ov\ee^2} \right)
$$
$$
= \ee^{4p^2 }
{12\ \Gamma ( -{3\ov 2}-2p^2)\ov\G ( {1\ov 2} )
\G (1-2p^2)}+
 {1\ov\ee^3}{12\ \Gamma ({3\ov 2} + 2p^2)\ov\G ({1\ov 2} )
\G (4+2p^2)}-{1\ov\ee}{45 \Gamma ({1\ov 2}+2p^2 )\ov \G ({1\ov 2})
\G (4+2p^2)}.
$$
The last two terms in \BBf\ are given by
$$
B_{\m\n} (p) B_{\rho\s} (-p)\tilde{K}^{\m\n\rho\s}(p)=
 B_{\m\n} (p)B^{\m\n}(-p){2(3+6p^2+4p^4)
\Gamma ( {3\ov 2} + 2p^2 )\ov
\G ( {1\ov 2} ) \G (4+2p^2)}
$$
$$ - p^\m p^\n B_{\m\n} (p)B^{\rho}_{\rho}(-p) {4\
\Gamma ( {1\ov 2} + 2p^2 )\ov \G ( {1\ov 2} )
\G (2+2p^2)}
$$
$$
+ {12\ \Gamma ( {3\ov 2} + 2p^2 )\ov \G ( {1\ov 2} )
\G (4+2p^2)}\left( p^\m p^\n B_{\m\n} (p) p^\rho p^\s B_{\rho\s} (-p) -
{4\ov 3}p^2 p^\n p^\s B_{\m\n} (p) B_{\m\s} (-p) \right) ,
$$
$$
B_{\m\n} (p) B_{\rho\s} (-p)K^{\m\n\rho\s}(p)=
 -B_{\m\n} (p)B^{\m\n}(-p){8p^2
\Gamma ( {3\ov 2} + 2p^2 )\ov
\G ( {1\ov 2} ) \G (4+2p^2)}
$$
$$ + p^\m p^\n B_{\m\n} (p)B^{\rho}_{\rho}(-p) {3\
\Gamma ( -{1\ov 2} + 2p^2 )\ov \G ( {1\ov 2} )
\G (2+2p^2)}
$$
$$
- {45\ \Gamma ( {1\ov 2} + 2p^2 )\ov \G ( {1\ov 2} )
\G (4+2p^2)}\left( p^\m p^\n B_{\m\n} (p) p^\rho p^\s B_{\rho\s} (-p) -
{4\ov 3}p^2 p^\n p^\s B_{\m\n} (p) B_{\m\s} (-p) \right)
$$

\listrefs

\end